\documentclass[usegraphicx]{mn2e}

\usepackage{times}
\usepackage{amssymb}
\usepackage{amsmath}

\begin{document}

\include{defn}
\def\cm{{\rm\thinspace cm}}
\def\gm{{\rm\thinspace gm}}
\def\dyn{{\rm\thinspace dyn}}
\def\erg{{\rm\thinspace erg}}
\def\eV{{\rm\thinspace eV}}
\def\MeV{{\rm\thinspace MeV}}
\def\g{{\rm\thinspace g}}
\def\ga{{\rm\thinspace gauss}}
\def\K{{\rm\thinspace K}}
\def\keV{{\rm\thinspace keV}}
\def\km{{\rm\thinspace km}}
\def\kpc{{\rm\thinspace kpc}}
\def\Lsun{\hbox{$\rm\thinspace L_{\odot}$}}
\def\m{{\rm\thinspace m}}
\def\Mpc{{\rm\thinspace Mpc}}
\def\Msun{\hbox{$\rm\thinspace M_{\odot}$}}
\def\Zsun{\hbox{$\rm\thinspace Z_{\odot}$}}
\def\pc{{\rm\thinspace pc}}
\def\ph{{\rm\thinspace ph}}
\def\s{{\rm\thinspace s}}
\def\yr{{\rm\thinspace yr}}
\def\sr{{\rm\thinspace sr}}
\def\Hz{{\rm\thinspace Hz}}
\def\MHz{{\rm\thinspace MHz}}
\def\GHz{{\rm\thinspace GHz}}
\def\chisq{\hbox{$\chi^2$}}
\def\delchi{\hbox{$\Delta\chi$}}
\def\cmps{\hbox{$\cm\s^{-1}\,$}}
\def\cmpssq{\hbox{$\cm\s^{-2}\,$}}
\def\cmsq{\hbox{$\cm^2\,$}}
\def\cmcu{\hbox{$\cm^3\,$}}
\def\pcmcu{\hbox{$\cm^{-3}\,$}}
\def\pcmcuK{\hbox{$\cm^{-3}\K\,$}}
\def\dynpcmsq{\hbox{$\dyn\cm^{-2}\,$}}
\def\ergcmcups{\hbox{$\erg\cm^3\ps\,$}}
\def\ergpcmps{\hbox{$\erg\cm^{-3}\s^{-1}\,$}}
\def\ergpcmsqps{\hbox{$\erg\cm^{-2}\s^{-1}\,$}}
\def\ergpcmsqpspA{\hbox{$\erg\cm^{-2}\s^{-1}$\AA$^{-1}\,$}}
\def\ergpcmsqpspsr{\hbox{$\erg\cm^{-2}\s^{-1}\sr^{-1}\,$}}
\def\ergpcmcups{\hbox{$\erg\cm^{-3}\s^{-1}\,$}}
\def\ergpcmps{\hbox{$\erg\cm^{-1}\s^{-1}$}}
\def\ergps{\hbox{$\erg\s^{-1}\,$}}
\def\ergpspmp{\hbox{$\erg\s^{-1}\Mpc^{-3}\,$}}
\def\gpcm{\hbox{$\g\cm^{-3}\,$}}
\def\gpcmps{\hbox{$\g\cm^{-3}\s^{-1}\,$}}
\def\gps{\hbox{$\g\s^{-1}\,$}}
\def\Jy{{\rm Jy}}
\def\keVpcmsqpspsr{\hbox{$\keV\cm^{-2}\s^{-1}\sr^{-1}\,$}}
\def\kmps{\hbox{$\km\s^{-1}\,$}}
\def\kmpspmp{\hbox{$\km\s^{-1}\Mpc{-1}\,$}}
\def\Lsunppc{\hbox{$\Lsun\pc^{-3}\,$}}
\def\Msunpc{\hbox{$\Msun\pc^{-3}\,$}}
\def\Msunpkpc{\hbox{$\Msun\kpc^{-1}\,$}}
\def\Msunppc{\hbox{$\Msun\pc^{-3}\,$}}
\def\Msunppcpyr{\hbox{$\Msun\pc^{-3}\yr^{-1}\,$}}
\def\Msunpyr{\hbox{$\Msun\yr^{-1}\,$}}
\def\pcm{\hbox{$\cm^{-3}\,$}}
\def\pcmsq{\hbox{$\cm^{-2}\,$}}
\def\pcmK{\hbox{$\cm^{-3}\K$}}
\def\phpcmsqps{\hbox{$\ph\cm^{-2}\s^{-1}\,$}}
\def\pHz{\hbox{$\Hz^{-1}\,$}}
\def\pmpc{\hbox{$\Mpc^{-1}\,$}}
\def\pmpccu{\hbox{$\Mpc^{-3}\,$}}
\def\ps{\hbox{$\s^{-1}\,$}}
\def\psqcm{\hbox{$\cm^{-2}\,$}}
\def\psr{\hbox{$\sr^{-1}\,$}}
\def\kmpspMpc{\hbox{$\kmps\Mpc^{-1}$}}

\voffset=-0.4in

\title{On viscosity, conduction and sound waves in the intracluster
medium}
\author[A.C. Fabian et al]
{\parbox[]{6.in} {A.C. Fabian$^1$, C.S. Reynolds$^2$, G.B. Taylor$^{3,4}$ and 
R.J.H.~Dunn$^1$\\
\footnotesize
1. Institute of Astronomy, Madingley Road, Cambridge CB3 0HA\\
2. Department of Astronomy, University of Maryland, College Park, MD
20742, USA \\
3. Kavli Institute of Particle Astrophysics and Cosmology, Menlo Park, 
CA 94025, USA \\
4. National Radio Astronomy Observatory, P.O. Box O, Socorro, NM
87801, USA
}}

\maketitle \begin{abstract} Recent X-ray and optical observations of
the Perseus cluster indicate that a combination of weak shocks at
small radii ($\gtrsim20\kpc$) and viscous and conductive dissipation of sound
waves at larger radii is responsible for heating the intracluster
medium and can balance radiative cooling of cluster cores. We
discuss this mechanism more generally and show how the specific
heating and cooling rates vary with temperature and radius. It appears
that this heating mechanism is most effective above $10^7\K$, which
allows for radiative cooling to proceed within normal galaxy formation
but stifles the growth of very massive galaxies.  The scaling of the
wavelength of sound waves with cluster temperature and feedback in the
system are investigated. \end{abstract} 

\section{Introduction}
The H$\alpha$-emitting filaments surrounding NGC\,1275 in the core of
the Perseus cluster (Lynds 1970; Conselice et al 2001) appear to act
as streamlines revealing flows in the intracluster medium (Fabian et
al 2003b). Many of the outer filaments are reasonably straight and
stretch over 10s kpc. One dubbed the horseshoe (Conselice et al 2001)
folds back on itself just behind a large spherical-cap shaped
depression in  the X-ray intensity of the gas, suggestive of the flow
on Earth seen behind rising spherical-cap bubbles in water (Fabian et
al 2003b). Together these observations indicate that the intracluster
medium is not highly turbulent and is probably viscous, with a viscosity
approaching the Spitzer-Braginskii (1956) value for an ionized plasma.

Further evidence for the medium being viscous is provided by the {\em
shape} of detached buoyant bubbles, seen most clearly in the Perseus
cluster (Fabian et al 2003a). Simulations of bubbles rising in an
intracluster medium show that viscosity enables bubbles to remain
intact for longer than a crossing time (Reynolds et al 2005).Strong
magnetic fields are otherwise required to prevent bubbles from breaking
up.

If the intracluster medium is viscous then sound waves produced
for example by radio bubbles from the central active galaxy, as seen
in the Perseus cluster (Fabian et al 2003a), can be rapidly
dissipated. This provides an efficient mechanism for transferring
energy produced by a massive black hole in a central cluster galaxy
into the surrounding medium in a reasonably isotropic manner (see also
Ruszkowski, Br\"uggen \& Begelman 2004a,b). Although
the energy flow passes through highly directional jets, the resultant
bubbling produces approximately spherical sound waves which propagate
into the cluster.

Energetic outflows from black holes do not otherwise couple well to
surrounding gas. Electromagnetic ones pass through the gas which,
being ionized, is mostly transparent. Collimated outflows, which tend
to be the most energetic, are often too violent and bore right through
the gas along the collimation direction (e.g. Cygnus A; Hydra~A,
McNamara et al 2005). Uncollimated winds might work but would tend to
give strong shocks, which are not observed in the inner intracluster
gas. Viscous dissipation can provide continuous, gentle, distributed
heat (Fabian et al 2003a), as is required (Voigt \& Fabian 2004).


Here, we examine the dissipation of sound wave energy in intracluster
and intragroup gas due to both viscosity and thermal conduction.  Both
of these transport processes have a strong temperature dependence
which means that such heating dominates over radiative cooling above
about $10^7\K$. Radiative cooling dominates below that temperature.
This provides a clue as to why heating appears to stifle the cooling
in clusters and groups (the cooling flow problem, see Fabian et al
2001) yet most galaxy formation, where much of the stellar component
is a consequence of radiative cooling of baryons which have fallen
into dark matter wells, has taken place. Indeed the process which
stifles cooling in massive systems probably is responsible for
truncating the upper mass range of galaxies (Fabian et al 2002; Benson
et al 2003; Binney 2004).

\section{Viscous and conductive heating}

\subsection{Sound wave heating and radiative cooling rates}


We now consider the ICM heating rate caused by the dissipation of
radio-galaxy induced sound waves.  Suppose the ICM possesses a
kinematic viscosity $\nu$ and a thermal conductivity of $\kappa$.  The
absorption coefficient for the passage of sounds waves is given by
(Landau \& Lifshitz 1987)
\begin{equation}
\gamma=\frac{2\pi^2f^2}{c_{\rm s}^3}\left[\frac{4}{3}\nu + \frac{\kappa}{\rho}\left(\frac{1}{c_{\rm V}}-\frac{1}{c_{\rm p}}\right)\right],
\end{equation}
where $f$ is the frequency of the sound waves, $c_{\rm s}$ is the
sound speed, $\rho$ is the density of the ICM, and $c_{\rm V}$ and
$c_{\rm p}$ are the specific heats at constant volume and pressure,
respectively.  The dissipation length (i.e., the length scale over
which the energy flux of the sound wave decrease by a factor of $e$)
is given by $\ell=1/2\gamma$.  Noting that $c_{\rm p}/c_{\rm
  V}=5/3$ for a fully ionized gas, we can write
\begin{equation}\label{eqn:dissipation_length}
\ell=\frac{3c_{\rm s}^3}{8\pi^2f^2\left[2\nu+\kappa/\rho c_{\rm p}\right]}.
\end{equation}
We note that this expression is only strictly true in the regime in
which the wavelength of the sound waves are short compared with the
dissipation length $\ell$.  We briefly discuss a possible violation of
this condition later in this Section.  Now, suppose that the viscosity
and thermal conductivity are fixed fractions, $\xi_{\nu}$ and
$\xi_{\kappa}$, of their unmagnetized values.  For a fully ionized
hydrogen plasma, in c.g.s. units, we get
\begin{eqnarray}
\nu&=&1.0\times 10^{25}T_7^{5/2}n^{-1}\xi_\nu,\\
\frac{\kappa}{\rho c_{\rm p}}&=&2.36\times 10^{26}T_7^{5/2}n^{-1}\xi_{\kappa},
\end{eqnarray}
where $T_7=T/(10^7\K)$, and $n$ is the electron number density, and we
have taken the Coulomb logarithm to have a value of 37.  Substituting
the numerical values into eqn.~\ref{eqn:dissipation_length}, we get
\begin{equation}
\ell=697\,\frac{nT_7^{-1}f_{-6}^{-2}}{\left(\frac{\xi_\nu}{0.1}\right)+11.8\left(\frac{\xi_\kappa}{0.1}\right)}\kpc
\end{equation}
where we have expressed the sound wave frequency in units of per
megayear (i.e., $f_{-6}=f/(10^{-6}\yr^{-1})$ and have normalized to
plausible values of viscosity and thermal conductivity (following the
arguments of Narayan \& Medvedev 2001). Hereafter, we shall denote
\begin{equation}
{\bar \xi}=\frac{\xi_{\nu}}{0.1}+11.8\frac{\xi_\kappa}{0.1}
\end{equation}
We note that, since temperature and density vary with radius in the
cluster, the dissipative effects of viscosity and thermal conductivity
vary but maintain a fixed ratio (as a consequence of assuming that
$\xi_\nu$ and $\xi_\kappa$ are constant).  We also note that thermal
conduction is over an order of magnitude more effective at dissipating
sound wave energy if $\xi_\nu\approx\xi_\kappa$.

We now model the central AGN as a source of acoustic energy at the
center of the cluster.  Suppose that the AGN injects an acoustic
luminosity of $L_{\rm inj}$ into the ICM at an inner radius of
$r=r_{\rm in}$.  If $L_s(r)$ is the acoustic luminosity at radius
$r>r_{\rm in}$, the definition of dissipation length gives
\begin{equation}
\frac{dL_s}{dr}=-\frac{L_s}{\ell},
\end{equation}
which has the solution
\begin{equation}
L_s(r)=L_{\rm inj}\exp\left(-\int_{r_{\rm in}}^r\frac{1}{\ell}\,dr\right).
\end{equation}
The volume heating rate due to viscous and conductive dissipation is
\begin{equation}
\epsilon_{\rm diss}=\frac{L_s(r)}{4\pi r^2\ell},
\end{equation}
which, for $r\ll\ell$ is approximately
\begin{equation}
\epsilon_{\rm diss}\approx\frac{2\pi f^2L_{\rm inj}\left[2\nu+\kappa/\rho c_{\rm p}\right]}{3 c_{\rm s}^3 r^2},
\end{equation}
and can be evaluated to give
\begin{equation}
\epsilon_{\rm diss}\approx3.8\times 10^{-29}T_7n^{-1}f_{-6}^2r_2^{-2}L_{44}{\bar \xi}\ergpcmcups,
\end{equation}
where $L_{44}=L_{\rm inj}/(10^{44}\ergps)$ and $r_2=r/(100\kpc)$.

\begin{figure}
\includegraphics[width=1.0\columnwidth]{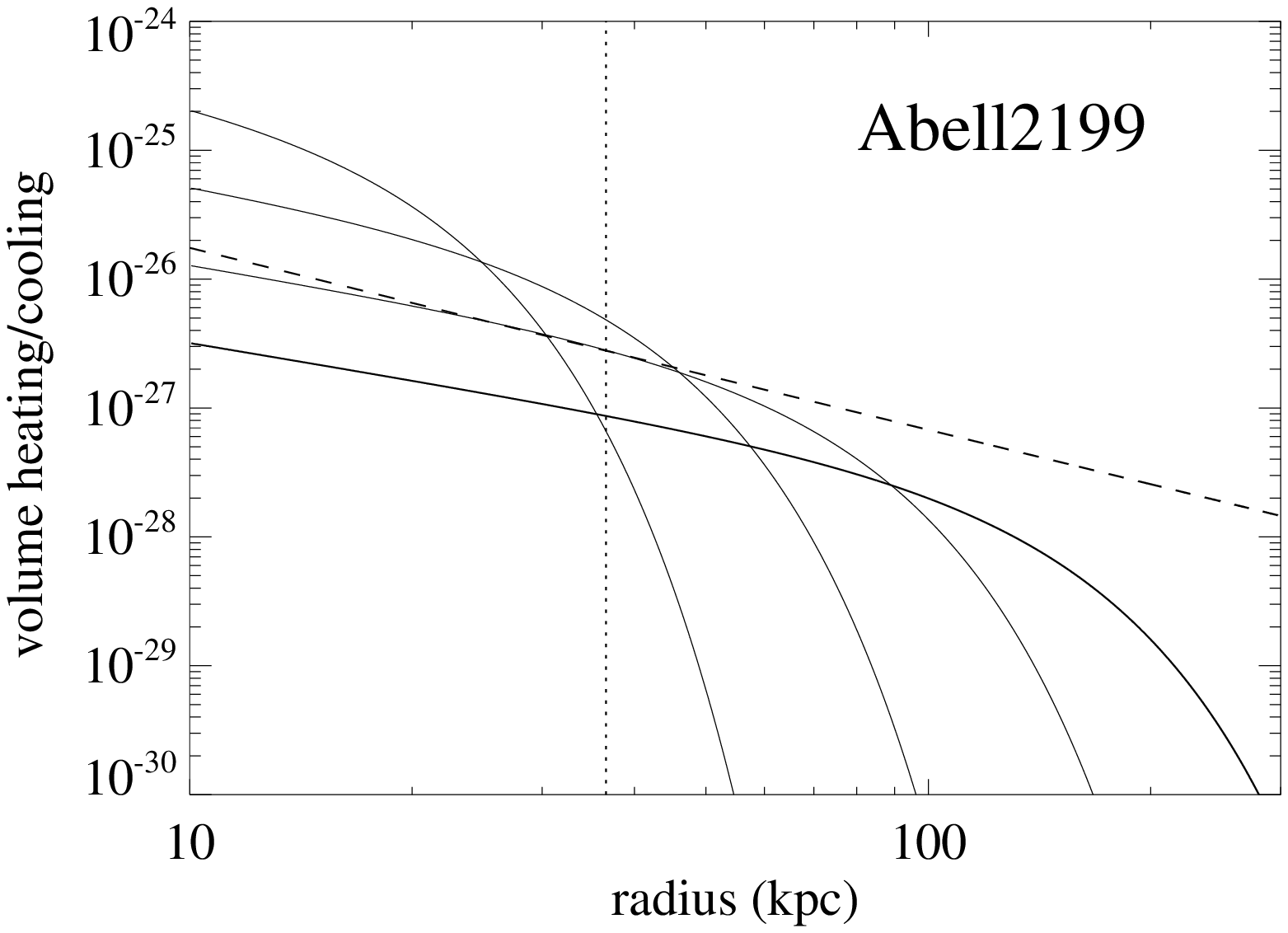}

\includegraphics[width=1.0\columnwidth]{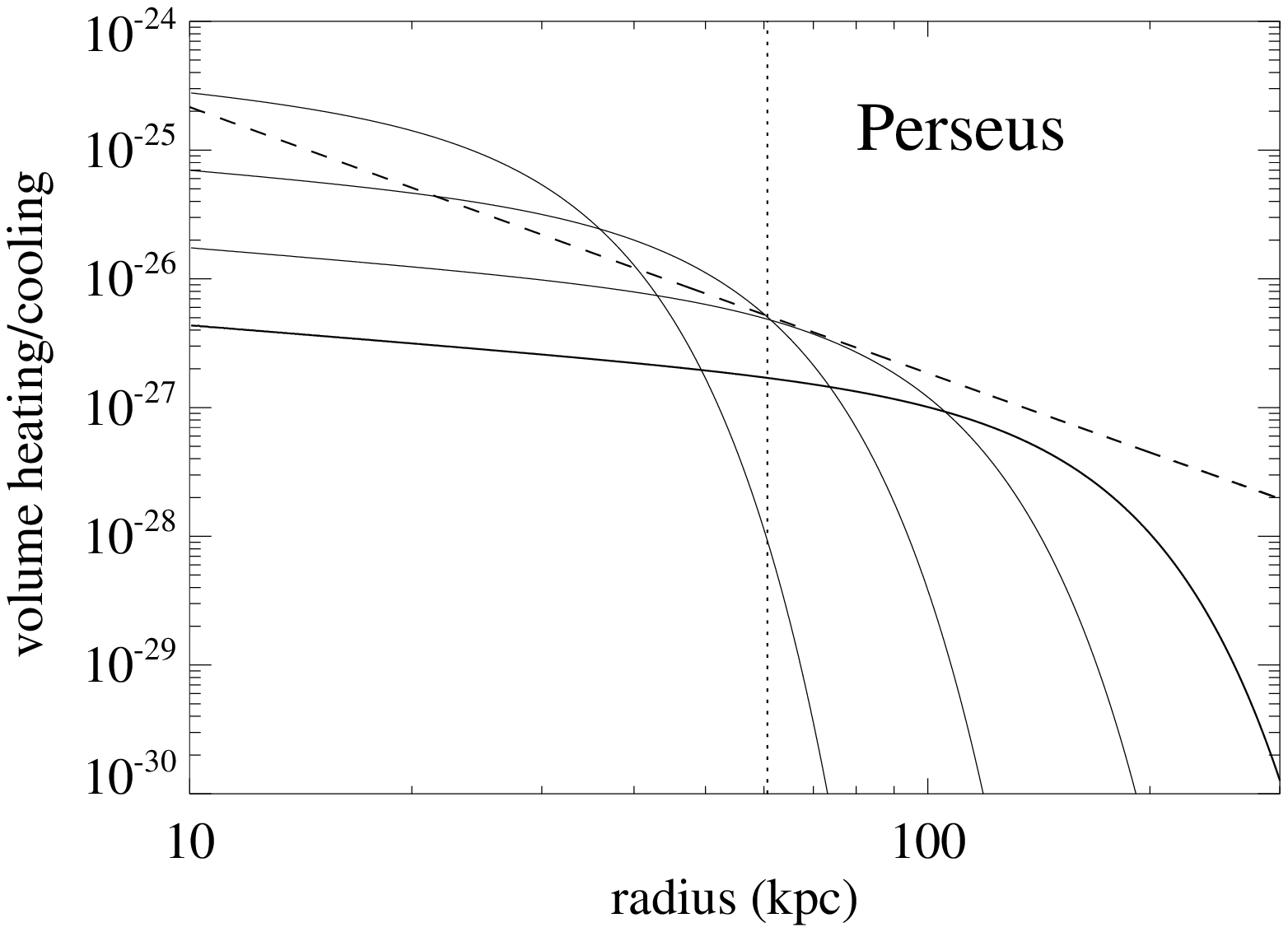}

\caption{Dissipative volume heating rates (solid lines) and radiative
volume cooling rate (dashed line), using the temperature and density
profile of (Top) A~2199 (Johnstone et al. 2002) and (Bottom) the core
of the Perseus cluster (Sanders et al 2004).  The heavy solid line is
for the following set of illustrative set of parameters for which the
heating and cooling approximately balance; $L_{44}=1$ for A\,2199, 5
for Perseus, $f_{-6}=0.1$, $\xi_\nu=0.1$, $\xi_\kappa=0$. Other
choices of the acoustic power ($L_{44}$ will simply rescale this line
in the vertical direction.  Other choices of $f_{-6}$, $\xi_\nu=0.1$,
and $\xi_\kappa=0$ change the dissipation lengthscale as well --- the
family of additional (thin) solid lines shows the effect of successive
doublings of frequency ($f_{-6}=0.2$, $f_{-6}=0.4$, and $f_{-6}=0.8$).
The vertical dashed lines indicate a fiducial cooling radius where the
radiative cooling time of the gas is 3~Gyr.} \end{figure}


As illustrations of the radial dependence of the viscous and
conductive dissipation, we compute $\epsilon_{\rm diss}$ using the
temperature and density profiles of Abell~2199 from Johnstone et al.
(2002; $T=4.4r_2^{0.3}\keV$ and $n=6.0\times
10^{-3}r_2^{-0.75}\pcmcu$) and beyond about 20~kpc in the core of the
Perseus cluster from Sanders et al (2004; $T=5.84r_2^{0.44}\keV$ and
$n=1.0\times 10^{-2}r_2^{-1.1} \pcmcu$).  We compare these with the
radiative cooling as derived from the formula of Tozzi \& Norman
(2001) which, expressed in our units, becomes
\begin{equation}
\epsilon_{\rm rad}=10^{-24}n_in_e\left[1.13T_7^{-1.7}+5.3T_7^{0.5}+6.3\right]
\end{equation}
As shown in Fig.~1, the dissipative heating can be comparable or
exceed the radiative cooling within the fiducial cooling radius in
which the cooling time $\tau=3nkT/2\epsilon_{\rm rad}$ exceeds 3\,Gyr.
Indeed, in the case of Abell~2199, the dissipative heating closely
balances the radiative cooling out to the cooling radius for the
following choice of parameters; $L_{44}=1$, $f=0.2$, $\xi_\nu=0.1$,
$\xi_\kappa=0$.  The dissipation lengthscale is approximately
100\,kpc, and the heating rate displays power-law radial profile
within this radius (reflecting the simple power-law forms of the
assumed temperature and density).  A sharp (exponential) cutoff in the
heating rate is seen for radii larger than the dissipation length.
Varying the acoustic luminosity of the AGN simply changes the
normalization of this heating rate.  On the other hand, varying the
frequency of the sound waves leads to dramatic changes in the
dissipation lengthscale.  In particular, increasing the sound
frequency while leaving the other parameters at the values listed
above rapidly brings the dissipation lengthscale within the cooling
radius.  A detailed balance of heating and cooling is then no longer
possible --- the innermost regions can be strongly heated while
radiative cooling occurs unchecked at the cooling radius.  In the case
of the Perseus cluster, the steep density profile results in the
radiative cooling possessing a significantly stronger radial
dependence than the ``monochromatic'' (i.e., single frequency)
dissipative heating profile.  Thus, any balance between sound wave
dissipation and radiative cooling must employ acoustic waves with a
range of frequencies.


To quantify this, suppose that the injected acoustic luminosity
between frequencies $f$ and $f+df$ is $L_{\rm inj}(f)\,df$. Furthermore,
suppose that all frequencies are injected at the same radius $r=r_{\rm
in}$.  The full expression for the dissipative heating is then,
\begin{equation}
\epsilon_{\rm diss}=\int_0^\infty\frac{L_{\rm inj}(f)}{4\pi r^2\ell}\exp\left( -\int_{r_{\rm in}}^r\frac{1}{\ell}\,dr\right)\,df.
\end{equation}
For comparison with Perseus, it is useful to quantify the flattening
of the dissipation profile due to the spectrum of frequencies in the
case of power-law radial dependences of temperature and density.
Suppose that $T\propto r^\alpha$ and $n\propto r^{-\beta}$, where we
expect $\alpha,\beta>0$ on general physical grounds.  Using eqn (11)
we can then approximate the dissipative heating profile of a given
frequency as a power-law in radius,
\begin{equation}
\epsilon_{\rm diss}(f)\propto r^{\alpha+\beta-2}f^2L_{\rm inj}(f)
\end{equation}
which is truncated at radius $r=\ell$.  Suppose that the spectrum of
frequencies has a power-law form, $L_{\rm inj}(f)\propto f^{-\gamma}$
between $f=f_{\rm min}$ and $f=f_{\rm max}$.  Noting that $\ell$
increases with decreasing frequency, we can delineate the following
regimes;
\begin{enumerate}
\item If $r>\ell(f_{\rm min})$, the radius under consideration is
outside of the overall dissipation region and the dissipative heating
profile is rapidly declining.
\item If $r<\ell(f_{\rm max})$, even the highest frequency component
has not yet dissipated away and the radial dependence of the heating
is just the sum of similar monochromatic components, $\epsilon_{\rm
diss}\propto
r^{\alpha+\beta-2}$.
\item If $\ell(f_{\rm max})<r<\ell(f_{\rm min})$, only the frequency components
upto $f=\tilde{f}$ contribute, where $\tilde{f}$ is defined by
$r=\ell(\tilde{f})$.  In this case,
\begin{equation}
\epsilon_{\rm diss}\propto \int_{f_{\rm
min}}^{\tilde{f}}r^{\alpha+\beta-2}f^{2-\gamma}\,df
\end{equation}
For $\gamma>3$, the lowest frequencies always dominate the energetics
and we again get the monochromatic case $\epsilon_{\rm diss}\propto
r^{\alpha+\beta-2}$.   For $\gamma<3$, we get $\epsilon_{\rm diss}\propto r^\delta$, where
\begin{equation}
\delta=\alpha+\beta-2 -\frac{(3-\gamma)(\alpha+\beta+1)}{2}.
\end{equation}
The second term on the RHS of the last expression represents the
modification of the monochromatic result due to the spectrum of
frequencies.
\end{enumerate}
Using these results, we can predict that for dissipative heating to
balance radiative cooling in the Perseus cluster, we require a
spectrum of fluctuations with $\gamma\approx 1.8$ across at least the
frequency range $f_{-6}=0.2-1$.


For a given suppression factor ($\xi_{\nu}$ and $\xi_{\kappa}$),
thermal conduction can be significantly more important than viscosity
at dissipating sound wave energy.  However, this is only strictly true
when the dissipation wavelength is large compared with the wavelength
of sound.  If the thermal conduction is such that the dissipation
lengthscale $\ell$ is comparable to or smaller than the wavelength of
the sound wave under consideration, dissipation due to thermal
conduction will be {\it ineffective} and the sounds wave will
propagate as an isothermal sound wave.  Indeed, if thermal conduction
occurs at the level suggested by Narayan \& Medvedev (2001),
\begin{equation} 
\xi_\kappa\approx 0.2,
\end{equation} 
the dissipation lengthscale would be extremely short
for all but the lowest frequency soundwaves ($\ell\approx
150f_{-6}^{-2}\pc$ using $n=0.02\pcmcu$ and $T_7=4$).  Thus an
implicit prediction of Narayan \& Medvedev (2001) is that the sound
waves observed in the Perseus cluster are isothermal sound waves.
Indeed, the thermal conduction can become so ineffective at
dissipating acoustic energy that viscous dissipation dominates even if
$\xi_{\kappa}> \xi_{\nu}$.

The theory outlined above only strictly applies to the case of linear
sound waves; this imposes a radius dependent maximum acoustic
luminosity to which this theory can be applied.  Expressed in terms of
pressure fluctuations, $\delta p$, the acoustic luminosity crossing a
shell at radius $r$ is,
\begin{equation}
L_s\approx 4\pi r^2\frac{\delta p^2}{\rho c_s}.
\end{equation}
The acoustic waves become strongly non-linear when $\delta p\approx
p$.  Thus, the maximum acoustic luminosity for which our model applies
is
\begin{equation}
L_{s,max}\approx 4\pi r^2 \frac{p^2}{\rho c_s}\approx 3.1\times 10^{46}nT_7^{1.5}r_2^2 \ergps.
\end{equation}
Using this equation, we see that the fiducial model described above
for A~2199 ($L_{44}=1$, $f=0.1$, $\xi_\nu=0.1$, $\xi_\kappa=0$) is
only valid for radii greater than about 20\,kpc.  Within that radius,
the energy must be transported within the form of shocks and/or strong
convective motions.

For comparison, we now consider the dissipation rate in a weak shock, 
$\epsilon_{\rm shock}\propto (\delta P/P)^3 p f$ (Landau \& Lifshitz
1987; Stein \& Schwartz 1972) 
\begin{equation}
\epsilon_{\rm shock}\approx 4\times 10^{-28} L_{44}^{3/2} n^{-1/2}
T_7^{-5/4} r_2^{-3} f_{-6} \ergpcmcups.
\end{equation}
This is comparable with the required heating rate at about 10--20~kpc radius in
the Perseus cluster (where a weak shock is seen, Fabian et al 2003a)
if $f_6\sim 0.1$ and $L_{44}\sim 10$. For the density and temperature profiles 
of that cluster, $\epsilon_{\rm diss}\propto r^{-0.46}$ and $\epsilon_{\rm
shock}\propto r^{-3}$. Viscous dissipation is therefore most
effective at larger radii and for higher frequencies, shock
dissipation applying most to small radii and low frequencies. Since 96
per cent of the {\em volume} within 60~kpc lies beyond 20~kpc, it is
reasonable to state that viscous heating can dominate the volume heating.

\subsection{A note on the effect of magnetic fields in the ICM}


In the above analysis, the viscosity and thermal conductivity were
assumed to be a fixed fraction of the Spitzer-Braginskii (1956) value
for an ionized gas.  Cluster cores do have significant magnetic fields
(e.g. Carilli \& Taylor 2002) which may of course alter the
viscosity. What effect this has is unclear.

There is, however, evidence from Faraday rotation measures (RMs)
observations (Carilli \& Taylor 2002 and references therein) that the
field is coherent in patches larger than the ion mean free path.
Typical values of the cell size estimated from RM distributions and
quoted in the literature are 5-10 kpc.  A power-spectrum analysis of
Hydra~A, a bright radio galaxy and one of the best cases for probing a
cooling core cluster as it provides a probe of the RM distribution on
scales from 0.2 to 40 kpc, has been carried out by Vogt \& Esslin
(2005), who find a magnetic autocorrelation length of
$3\pm0.5\kpc$.  This autocorrelation length can be compared to the ion
mean free path (mfp) of $0.23 T_{\rm c}^2 n_{\rm c}^{-1} \kpc$ for
typical cluster core parameters of $ T_{\rm c}=3\times 10^7\K$ and
$n_{\rm c}^{-1}= 0.01\pcmcu$.  Since the mfp is smaller than the
dominant scale size of the magnetic field, the viscosity along the
magnetic field direction will be high
for much of the volume under consideration. This is also relevant to
the issue of conduction, which can also dissipate the energy in sound
waves (Landau \& Lifshitz 1987; see also Ruszkowski, Br\"uggen \&
Hallman 2005), but is not strong enough to offset
cooling in many clusters by itself (Voigt \& Fabian 2004; Kaastra et
al 2004).

The simple model considered in Section 2.1 can be motivated from the
notion that intracluster medium consists of many magnetically isolated
blobs of gas, each of size ranging over a few kpc. Within a blob the
field is fairly coherent. Conduction of heat from the outside of the
core to the centre is inhibited by the boundaries between the
blobs. The energy in sound waves propagates freely across the blob
boundaries and can be dissipated within the blobs by both conduction
and viscosity.

There are other possible sources of viscosity in a tangled magnetic
field. For example if the field is in the form of flux tubes there can
be drag between the high field regions and the low field ones
(Longcope, McLeish \& Fisher 2003). Alternatively, the spatial
configuration of the field may change as sound waves pass across a
region in the sense that shaking up the field causes it to shift to a
lower energy configuration, so releasing energy either immediately or
later. This would operate as a bulk viscosity. (The issue of bulk
viscosity is also mentioned by Ruszkowski et al 2004). 

\section{The wavelength and frequency of sound waves}


Subcluster mergers are common in clusters and these will generate
large pressure discontinuities and sound waves. The wavelengths of
sound waves remaining after shocks have passed will presumably be on
the scale of cluster cores or hundreds of kpc. On these scales the gas
may be turbulent so the motions may degrade down to scales of tens of
kpc where viscous dissipation can operate. The outer gas in a cluster
may therefore be noisy with a wide spectrum of wavelengths.  Pringle
(1989) noted that the dense regions of a cluster core can focus sound
waves inward, meaning that significant heating may result at the
centre. Fujita et al (2004), using a 2D simulation, claim that such
motions generate strong turbulence. Establishing the energetics and
efficiency of such processes requires numerical calculations beyond
the scope of this paper. We note that propagating turbulence inward,
where the gravitational potential is steep and the gas dense, is
energetically challenging.

A central active nucleus giving rise to jets will inflate cavities of
relativistic radio-emitting plasma within the intracluster medium.
Such cavities or bubbles are seen in the Perseus cluster (B\"ohringer
et al 1993; Fabian et al 2000; Fabian et al 2003a), Virgo cluster
(Wilson et al 2002; Forman et al 2003), A2052 (Blanton et al 2001, 2003),
A2597 (Nulsen et al 2002), A4059 (Heinz et al 2002) and Hydra A
(McNamara et al 2000). Such bubbles inflate and then break away due to
buoyancy (Churazov et al 2000, 2001) and so even a constant jet will
lead to a bubbling effect. This generates a repetitive pressure excess
and thus a sound wave with a period equal to the repetition time of
the bubbles.

A bubble expands because of its excess pressure relative to the
intracluster medium at a rate which varies as $t^{1/3}$, where $t$ is
time. It detaches from the jet when buoyancy dominates which is 
roughly when its expansion velocity drops to about one half of the
local Keplerian velocity due to the local gravitational field of the
central galaxy and cluster, $v_{\rm K}$ (Churazov et al 2000). A new
bubble then begins to grow. This gives a natural bubbling period $f
\propto L^{-1/2} P^{1/2} v_{\rm K}^{3/2}$, where $P$ is the pressure
of the intracluster gas and $L$ is the power of the jet. Now the
central cooling time of many clusters with cool cores levels off at
about $10^8\yr$ within a radius of 10~kpc (Voigt \& Fabian 2004) which
implies that (between clusters) $P\propto T^{3/2}$.  If we also assume
that $v_{\rm K}\propto T^{1/2}$ where the gas temperature is $T$. 
Then $f \propto L^{-1/2} T^{3/2}$ and further scaling depends on $L$.

The dependence of $L$ on $T$ is probably fairly steep. If we assume
that the jet power $L$ does balance cooling then it is proportional to
the X-ray luminosity $L_{\rm cool}$, within the cooling radius (where
the cooling time equals the cluster age). $L_{\rm cool}$ is tabulated
for nearby clusters by Peres et al (1998) and using these values and
the cluster temperatures we find (Fig.~2) that $L\propto T^{\alpha}$
with $\alpha\sim 4$. Therefore $f \propto T^{-1/2}$. As a rough check,
the scaling predicts that radius $R$ when a bubble separates,
$R\propto L^{1/2}P^{-1/2}v_{\rm K}^{-1/2}\propto T$.  This scaling
result is shown in Fig.~3 with results for individual clusters drawn
from the compilation of Dunn \& Fabian (2004). 

\begin{figure} 
\includegraphics[angle=-90,width=0.95\columnwidth]{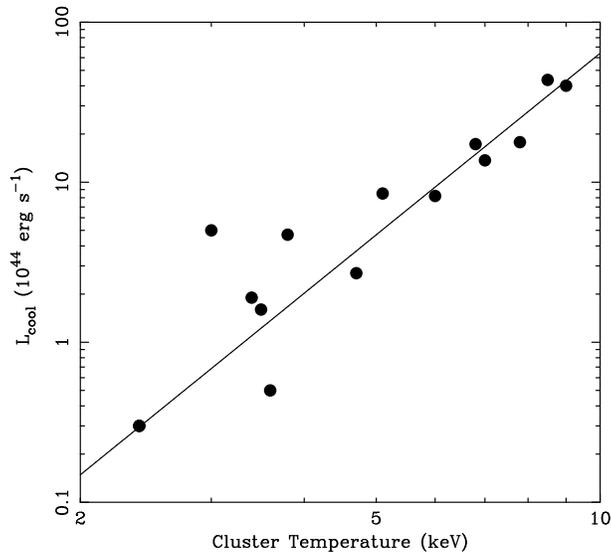} 
\caption{The X-ray luminosity within the cooling radius plotted
against the cluster temperature, using results for cooling flow
clusters taken from Peres et al (1998). The best fitting power law is
shown, with index 3.8.  } 
\end{figure}

\begin{figure} 
\includegraphics[angle=-90,width=0.95\columnwidth]{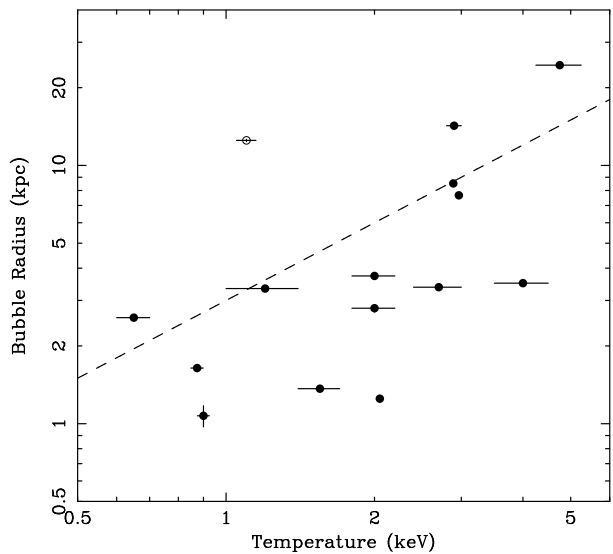} 
\caption{The radius of bubbles is shown
as a function of the surrounding temperature. Data are for attached
bubbles from the compilation of Dunn \& Fabian (2004). The open point
is for A2052 where there appears to be much cool gas accumulated
around the rims (Blanton et al 2003) so the points may reasonably
shift to the right. The dashed line is not a fit but indicates the
$R\propto T$ relation tentatively derived in the text for the rough
radius, $R$, at which bubbles detach from the nucleus. Bubbles should
therefore only exist below, or close to, the line.  } 
\end{figure}

The luminosity drops out of the dissipation formula (11) which now
varies steeply with temperature; $\epsilon_{\rm diss}\propto
T^{7/2}r^{-2}$ and at a given radius $\epsilon_{\rm
diss}/\epsilon_{\rm rad}\propto T^2$. 

Although we have in this Section identified the bubbling frequency as
the dominant one, we speculate that the underlying variability of the
radio jets occurs at higher frequencies and carries much power.  Most
jets are not constant as shown by repeated outbursts on small scales
(e.g., in Perseus A=3C84, Taylor \& Vermeulen 1996, see also Reynolds
\& Begelman 1997, and in general Zensus 1997).  These variations will
be transmitted through to the surface of the bubble leading to its
growth being erratic and a spectrum of high frequency sound waves
being present at small radii, with the bubbling frequency being the
low-frequency cutoff. The steep dependence on frequency means that
higher frequency waves are damped most easily. A simple estimate of
the heating effect of bubbles from the bubble size and recurrence rate
(e.g. Birzan et al 2004) will underestimate the true heating rate.

\begin{figure} 
\includegraphics[angle=-90,width=0.95\columnwidth]{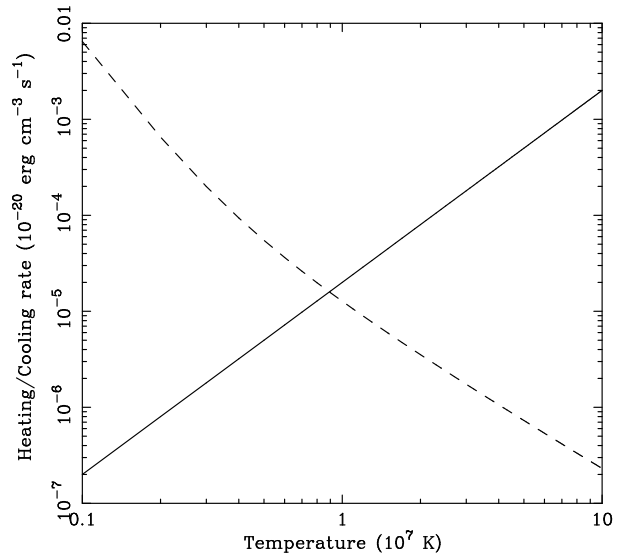} 
\caption{Heating (solid) and cooling (dotted) rates plotted as a function of
temperature for a pressure $nT=10^6\pcmcuK$ and radius of 10~kpc. For
the heating rate, $f_{-6}^2 L_{44}{\bar \xi}=1$.  }
\end{figure}

As an indication of the temperature dependence of the dissipation
versus radiative cooling we show $\epsilon_{\rm diss}$ and
$\epsilon_{\rm rad}$ as functions of temperature in Fig.~4. We assume
that $f_{-6}^2L_{44}{\bar \xi}=1, r=10\kpc$ and $P=nT=10^6\pcmcuK$
which means that if the above scaling is correct, the heating rate in
real objects is steeper. The plot shows that sound wave dissipative
heating dominates above about $10^7\K$ and radiation dominates
below. The process is therefore relevant as an explanation for the
truncation of cooling, and thus of star formation and the total
stellar mass, in massive galaxies. 

\section{Discussion}

We have shown that viscous heating of intracluster or intragroup gas
can effectively compete with radiative cooling above gas temperatures
of about $10^7\K$.  The repetitive $PdV$ work done by the formation of
bubbles is dissipated by weak inner shocks and outer viscous
processes. More rapid intrinsic variability in the jets leads to
quasi-continuous generation of higher frequency sound waves which
dissipate throughout the cooling region. Sound waves produced by jets
from a central black hole can thereby strongly reduce  cooling in the
surrounding gas and prevent further growth of the central galaxy by
accretion of radiatively cooled gas. This mechanism may explain the
upper-mass cutoff in galaxies since it operates best at high
temperatures.

If cooling initially dominated in a cluster core then the temperature
profile would have dropped to the {\em local} virial one, which for a NFW
potential (Navarro, Frenk \& White 1997) approaches $T\propto r^{1/2}$
for the central cusp (i.e. within the central 100~kpc).  Steady
heating then applied to this gas will tend to push the gas outward,
lowering its density whereas continued cooling means that gas flows
inward, in both circumstances following close to the local virial
temperature profile. This roughly explains the observed temperature
profiles, given that much of the inner few kpc are occupied by
bubbles. 

The prevalence of the central X-ray surface brightness peak
characteristic of short radiative cooling times (Peres et al 1998;
Bauer et al 2005) and the remarkable similarity of the cooling time
profiles in many clusters (Voigt \& Fabian 2004) suggests that a good
heating / cooling balance is commonly established.  This presumably
requires feedback with the central power source, the accreting black
hole. How that is established so that the properties of the gas at say
50~kpc are attuned to those at the black hole accretion radius of say
100~pc, is not completely clear.  

A partial answer can be seen from the following. If the radio source
becomes too vigorous, then most of its power will be deposited at
large radii beyond the cooling radius both if it becomes an FRII (as in
Cygnus A) or remains an FRI when the associated giant bubbles (as in
Hydra A, Nulsen et al 2004, or MS0735.6+7421, McNamara et al 2004)
create low frequency global disturbances which dissipate poorly. If
the radio source shuts off due to lack of fuel, then the inner gas
will cool and, angular momentum permitting, accrete and provide that
fuel. Rapid variations in jet power will lead to high frequency
disturbances which dissipate and so rapidly heat those inner regions.
A trickle of cooling and accretion explains the prevalence of warm
optical nebulosity (Crawford et al 1999 and references therein) and
cold gas (Edge 1991) near the centres of these objects. The inner
cooling time of these objects at $\sim 10^8\yr$ (Voigt \& Fabian 2004)
is similar to the dynamical timescale of the central galaxy.

Continued heating by a central AGN in cluster and group cores means
that the central black hole continues to grow. The total accreted mass
can be considerable (see Fabian et al 2002; Fujita \& Reiprich 2004).
This may skew the mass -- velocity dispersion relation ($M-\sigma$)
for massive black holes upward at the highest black hole masses, such
as expected of the massive black holes in groups and clusters. 

Finally, we note that a viscous intracluster medium means that motions
in the cores of relaxed clusters will be damped and, apart from the
regions directly around the inflating and rising bubbles, or where the
dark matter potential or galaxy are sloshing around, fluid motions
should be small and highly subsonic ($\lesssim0.3 c_{\rm s}$). Of relevance
here is the overall scale of any turbulent energy cascade, i.e. the
difference in length scale between the energy injection scale,
presumably the bubble size, and the dissipation scale, which may be as
large as the size of magnetic cells or similar distinct structures or
extend down to much smaller scales. Measurements of the widths and
shape of iron-K emission-lines with Suzaku will help to distinguish
laminar motions from turbulence (see Inogamov \& Sunyaev 2003 for
predictions of turbulent profiles).

\section{Acknowledgments}
We thank a referee for comments on shock dissipation.
CSR gratefully acknowledges support from the {\it Chandra} Cycle-5
Theory \& Modelling program under grant TM4-5007X. ACF thanks the
Royal Society for support.

\end{document}